\documentclass[a4paper]{article}
\usepackage{graphicx}
\title{The oscillations in the lossy medium}
\author{{\large Korotchenko K.B.}\\
{\small\bf Chair of Theoretical and Experimental Physics,}\\
{\small\bf Tomsk Polytechnical University, Rossia}}
\date{}

\begin{document}

\maketitle
\begin{abstract}
\noindent {\it The object of the work}~: to explore
dependence mass point oscillatory motion parameters in the
following cases:
\begin {itemize}
\item [\ -] without resistance (free oscillations);
\item [\ -] the resistance force is proportional to the velocity vector;
\item [\ -] the resistance force is proportional to the velocity squared.
\end {itemize}
{\it Used equipment}: the work have doing on a personal
computer. The oscillatory motion simulation is carried out by
the numerical solution of system of differential equations.
This equations describe a motion of a particle under an
elastic force action and exterior forces (resistance force)
with initial values and parameters being entered during the
dialogue with the computer.
\end{abstract}

\section{The theoretical part}

Let's begin from the {\it definition}: If some physical
quantity $F $ under specified physical conditions is
described {\bf periodic} or {\bf almost-periodic} function of
time one can say that this physical quantity is in
oscillatory process or {\bf in oscillations}.

As is known, a function $F (t) $ is called periodic if $F (t)
= F (t + T) $.\footnote{\small The definition almost-periodic
functions will be introduced later.} At the oscillatory
process the constant $T = 2\: \pi/\omega $ is called an {\bf
oscillation period} and the constant $\omega $ is called an
{\bf oscillation frequency (circular or cyclic)}. Obviously
$T $ is a time interval by means of that the values of
function $F(t) $ are repeated.

If the physical quantity is in oscillations described by the
{\it harmonic} function of time (i.e. function $\sin(\omega
t) $ or $\cos(\omega t) $) the oscillations is called {\bf
harmonic}.

Among all oscillatory processes the special interest is
represented those which the man can observe directly without
any devices. The most known oscillatory process having so
remarkable property is {\it the oscillatory motion}.

According to this, the {\bf oscillatory motion} of a mass
point we will call any its motion that the {\bf all} physical
quantities describing motion are {\bf periodic} (or {\bf
almost-periodic}) functions of time.

The major physical values describing a motion of a mass point
are:
\begin{itemize}
\item[\ -] the radius vector of a particle $\vec{r}(t) $, i.e.
its coordinates (we shall remind the equation of a form
$\vec{r} = \vec{r}(t) $ is called {\bf the motion equation}
(or {\bf law});
\item[\ -] and the vector of the particle acceleration $\vec{a}(t) $.
\end{itemize}
If we take into account that the vectors of velocity and
acceleration are defined {\it uniquely}  by the radius vector
$\vec{r} (t) $ of a mass point, it is possible to formulate
the following definition:
\begin{center}
\begin{tabular}{||c||}
\hline
any motion of a mass point\\
at which the radius vector of a particle\\
is a periodic (or almost periodic) function of time\\
is called the  {\bf oscillatory motion}\\
\hline
\end{tabular}
\end{center}

\subsection{Free simple harmonic motions}
\label{sb1}

The elementary oscillatory motion of a mass point is the {\it
harmonic} oscillatory motion. Thus, according to the definition
of harmonic oscillations, we shall called by {\bf free simple
harmonic} motion such oscillatory motion, at which the radius
vector of a particle is {\bf harmonic} function of time. It
means, the equation (the law) motion of a mass point that is
in a free harmonic oscillatory motion, has a form
\begin{equation}
\label{e1}
 \vec{r}(t) = \vec{r}_o \sin(\omega t + \varphi_o) .
\end{equation}
In eq.(\ref{e1}) the constant $\Phi = \omega t + \varphi_o $
is called by a {\bf phase} of the oscillatory motion and its
value at $t = 0 $, i.e. $\varphi_o\: $, is called by an {\bf
epoch angle} accordingly. The constant $\vec{r}_o $ is called
by an {\bf amplitude} of the oscillatory motion. From the
equation (\ref{e1}) is obvious that the amplitude is the
maximal value of radius vector the achievable at those point
in time when $\sin(\omega t + \varphi_o) = 1 $.

Let's note one important characteristic of the oscillatory
motion described by the equation (\ref {e1}). The vector
$\vec{r}_o $ is a {\it constant vector}, i.e. does not change
neither in magnitude nor in the direction. Therefore the
vector $ \vec{r}(t) $ can change {\it only} in magnitude (at
the expense of function $\sin(...) $), but remains parallel
to the {\it same} line. It means that {\it the harmonic
oscillatory motion always has only} {\bf one} {\it degree of
freedom}. In other words, {\it one coordinate is enough for
describing of a harmonic oscillatory motion}. For example,
coordinates measured along an axis $OX $. So the vector
equation (\ref{e1}) can always be replaced by one equation in
the coordinate form
\begin{equation}
\label{e1-1}
 x(t) = x_o \sin(\omega t + \varphi_o) ,
\end{equation}
where $x_o = |\vec{r}_o | $ is the module of the vector
$\vec{r}_o $.

It is easy to see that the equation (\ref {e1}) is {\it the
solution of the differential equation}
\begin{equation}
\label{e2}
 \frac{d^2 \vec{r}}{d t^2} + \omega^2 \vec{r} = 0\: .
\end{equation}
For this reason the differential equation (\ref{e2}) is
called by {\bf the equation of free simple harmonic motions}.
So, one can say that
\begin{center}
\begin{tabular}{||c||}
\hline {\bf free harmonic oscillatory motion}\\ of a mass
point is any motion described by\\ {\it the equation of free
simple harmonic motions} (eq.(\ref{e2}))\\ \hline
\end{tabular}
\end{center}
Classical example of a free harmonic oscillatory motion is
the particle motion with the mass $m $ due to action of
quasi-elastic force (i.e. simulative elastic force) $\vec{F}
= - k \vec{r} $, where $k $ is stiffness coefficient. To be
convinced of it we shall describe for such a mass point the
dynamical equation (i.e. Newton's second law)
\begin{equation}
\label{e3}
 m \vec{a} = - k \vec{r} .
\end{equation}
Taking into account that the acceleration is a second-order
derivative of the particle radius vector, we shall obtain
\begin{equation}
\label{e4}
 \frac{d^2 \vec{r}}{d t^2} + \frac{k}{m} \vec{r} = 0\: .
\end{equation}
Comparing the obtained equation with the equation of free
simple harmonic motions (\ref {e2}), we can see that the
motion a mass point due to action of quasi-elastic force is
really a free harmonic oscillatory motion. And the
oscillation cyclic frequency of a mass point is equal
\begin{equation}
\label{e5}
 \omega = \sqrt{\frac{k}{m}} .
\end{equation}

\subsection{Damped oscillations}
\label{sb2}

In the previous section we have considered a free harmonic
motion and were convinced that due to action of {\it only}
elastic force the mass point makes just such motion.

\begin{figure}
\includegraphics[width=5.5cm,height=3.5cm]{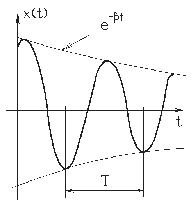}
\hfill
\includegraphics[width=5.5cm,height=3.5cm]{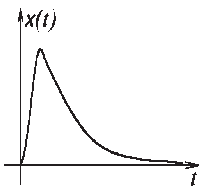}
\\
\parbox[t]{5.5cm}{\caption{Damped oscillation}\label{f1}}
\hfill
\parbox[t]{5.5cm}{\caption{Aperiodic oscillation}\label{f2}}
\end{figure}

Let's consider now motion a mass point due to action of
quasi-elastic forces $\vec{F} = - k \vec{r} $ in medium under
the action of {\it resistance forces}. Let, for example, the
resistance force is proportional to a vector of the particle
velocity $\vec{F_c} = - b \vec{v} $, where $b $ is the
resistance coefficient. Then the dynamical law (Newton's
second law) for such the mass point will have a form
\begin{equation}
\label{e6}
 m \vec{a} = \vec{F} + \vec{F_c} = - k \vec{r} - b \vec{v} .
\end{equation}
Taking into account that the velocity is a first-order
derivative and that the acceleration is a second-order
derivative of the particle radius vector, we shall obtain
\begin{equation}
\label{e7}
 \frac{d^2 \vec{r}}{d t^2} + \frac{b}{m} \frac{d\vec{r}}{d t}
                           + \frac{k}{m} \vec{r} = 0\: .
\end{equation}
It is easy to be convinced that the obtained equation
coincides with the equation of free simple harmonic motions
only at absence of the resistance forces (i.e. at $b = 0 $ ).
The solution of the equation (\ref{e7}) varies from the
solution of the equation (\ref{e2}) as well. The
eq.(\ref{e2}) is the equation of free simple harmonic
motions. So, the common solution of the equation (\ref{e7})
will have a form
\begin{equation}
\label{e8}
 \vec{r}(t) = \vec{r}_o e^{- \beta t} \sin(\omega t + \varphi_o)\: ,
\end{equation}
where the following notation for {\bf parameters of an
oscillatory motion} described by the equation (\ref{e7}) are
conventional
\begin{eqnarray}
\label{eqn01}
 \parbox[b]{8 cm}{{\bf damping factor}~~~} & - &
 \beta = \displaystyle\frac{b}{2 m} \nonumber \\
 \parbox[b]{8 cm}{
oscillation cyclic frequency of the free harmonic oscillatory
motion (i.e. at absence of the resistance forces)~~~} & - &
 \omega_o = \sqrt{\displaystyle\frac{k}{m}} \\
 \parbox[b]{8 cm}{
oscillation cyclic frequency of the studied harmonic
oscillatory motion~~~}                               & - &
 \omega = \sqrt{\omega_o - \beta^2}         \nonumber
\end{eqnarray}
Let's note that in these notation the equation (\ref {e7})
will look like
\begin{equation}
\label{e9}
 \frac{d^2 \vec{r}}{d t^2} + 2 \beta \frac{d\vec{r}}{d t}
                           + \omega_o \vec{r} = 0\: .
\end{equation}
As well as in the case of free simple harmonic
motions the oscillatory motion described by the equation
(\ref{e8}) has {\bf only one degree of freedom}. Hence, if to
set the direction of {\it constant} vector $\vec{r}_o $
parallelly to axis $OX $ of a cartesian frame, the
eq.(\ref{e8}) will have a form
\begin{equation}
\label{e8-1}
 x(t) = x_o e^{- \beta t} \sin(\omega t + \varphi_o)\: ,
\end{equation}
where, as well as in the equation (\ref{e1-1}), x is the
length of a vector $\vec{r}_o $.\\ In fig.\ref{f1} the qualitative
view of the solution (\ref{e8-1}) is presented. This figure
demonstrate that the studied oscillatory motion represents
oscillations with amplitude {\it decreasing in time} by
exponential law (i.e. described by the function $e^{- \beta
t} $). Just for this reason an oscillatory motion described
by the equation (\ref{e9}), named as {\bf the damped
oscillatory motion}. Accordingly, eq.(\ref{e9}) named as {\bf
the equation of damped oscillations}. So
\begin{center}
\begin{tabular}{||c||}
\hline {\bf the damped oscillatory motion}\\ of a mass
point\\ is any motion described by\\ the equation of damped
oscillations (i.e. eq.(\ref{e9}))\\ \hline
\end{tabular}
\end{center}
Let's consider more in detail properties of the damped
oscillatory motion. First of all it is obvious that in
contrast to the free harmonic oscillatory motion the radius
vector of the mass point in damped oscillations (i.e.
expression (\ref{e8}) or (\ref{e8-1})) is not periodic
function of time $\vec{r}(t) \neq \vec{r}(t+T) $. Thus {\it
damped oscillations} {\bf are not} {\it harmonic
oscillations}.

According to the definition by H. Bohr (Danish mathematician)
the function $f(t) $ satisfying the requirement
\begin{equation}
\label{e10}
 | f(t+T) - f(t) | < \epsilon ,
\end{equation}
where $\epsilon $ is some positive number is named an {\bf
almost-periodic} function. Accordingly, $T $ is named an {\bf
almost-period} such function. And the {\bf mean value} of an
{\it almost-periodic} function is always limitary
\begin{equation}
\label{e11} \lim\limits_{T \to \infty} \frac{1}{T}
\int\limits_{0}^{T} f(t) dt < \infty.
\end{equation}
It is easy to be convinced that for $x(t) $ from expression
(\ref{e8-1})
\begin{equation}
\label{e12} \lim\limits_{T \to \infty} \frac{1}{T}
\int\limits_{0}^{T} x(t) dt = 0 .
\end{equation}
Moreover, it always is possible to select such positive
number $\epsilon $ that the absolute value of the difference
$| x(t+T) - x(t) | $ (where $T = 2 \pi/\omega $) will be less
than this number. So the requirement (\ref{e10}) will be
satisfied.\\ Hence {\it radius vector of the mass point
making} the {\bf damped oscillations} is an {\bf
almost-periodic} {\it function with} {\bf almost-period} $T
$.

Let's remind that according to (\ref{eqn01}) the damped
oscillation cyclic frequency $\omega $ of mass point is equal
to
\begin{equation}
\label{e13}
 \omega = \sqrt{{\omega_o}^2 - \beta^2} .
\end{equation}
Obviously the quantity $\omega $ has the meaning of
oscillation {\it frequency} only in the case ${\omega_o}^2 <
\beta^2 $. At ${\omega_o}^2 > \beta^2\: $ the $\omega $
becomes {\it imaginary} and, accordingly, the trigonometrical
function $\sin(\omega t) $ is transformed to the hyperbolic
function $sh(\omega t) $. In this case the solution of the
damped oscillations equation (\ref{e9}) becomes
\begin{equation}
\label{e14}
 \vec{r}(t) = \vec{r}_o e^{- \beta t} sh(\omega t + \varphi_o)\: ,
\end{equation}
or in the coordinate notation
\begin{equation}
\label{e14-1}
 x(t) = x_o e^{- \beta t} sh(\omega t + \varphi_o)\: ,
\end{equation}
Such a solution is {\bf{\large neither}} a {\it periodic
function} {\bf{\large no}} an {\it almost-periodic function}.
And, therefore, the motion described by the equation of
damped oscillations at ${\omega_o}^2 > \beta^2 $ is
{\bf{\large not}} an oscillatory motion. This process is
named as {\bf aperiodic oscillations}. The diagram of a
function $x(t) $ for an aperiodic process (i.e. described by
eq.(\ref{e14-1}) at $\varphi_o = 0 $) is presented on fig.\ref{f2}.
\section{The practices for simulation of physical processes}

{\bf Before} simulation initiation of physical processes it
is {\bf necessary to familiarize} with {\it blanket rules of
operation with the digital computer and simple set of usual
activities} used at operation with the Borland software menu.
{\bf
\begin{center}
\begin{tabular}{||c||}
\hline In this laboratory work there is an opportunity\\ to
get the help information on its problem (and another)\\ at
any moment not quitting the program.\\ \hline
\end{tabular}
\end{center}}
To obtain the help information it is necessary to press the
key \fbox{$F1 $}.

The set of practices performed by the student at the study of
oscillatory motions are determined by the teacher and can
vary over a wide range.

Let's consider practices the realization of which is {\bf
necessary} for understanding of features of the mass point
oscillatory motion at presence (and absence) of resistance
forces. According to {\it the object of the laboratory work}
there should be two such practices.

\subsection{Free simple harmonic motions}
\label{sb4}

In this practice state problem to study an oscillatory motion
of a mass point at absence of resistance force. Namely:
\begin{itemize}
\item[\ -]
to make sure that a trajectory of a mass point is the
harmonic function;
\item[\ -]
to find out how the mass point trajectory varies by change of
the following parameters:
\begin{itemize}
\item[\ -]
particle mass $m $ and stiffness coefficient $k\: $ in
expression for elastic force (eq. (\ref{e3}))
\item[\ -]
initial kinematic parameters of a motion: the mass point
coordinates of the origin $x(0) = x_o \sin(\varphi_o) $ and
its initial velocity $v(0) = x_o \omega \cos(\varphi_o)\: $
(they are those parameters, you can change by changing value
of the oscillations epoch angle $\varphi_o $);
\end{itemize}
\end{itemize}
\noindent The practice consists of the following items:
\subsubsection{}
\label{ss1} After you entered in the menu and selected
necessary laboratory work (i.e. "The oscillatory motion") the
title page of this work arises.\\ Press \fbox{$Enter $} then
the {\it main} menu with titles of all practices will arise.
By keys \fbox{$\uparrow $} and \fbox{$\downarrow $} it is
necessary to select practice  "Simple harmonic Motions" and
press \fbox{$Enter $}.
\subsubsection{}
\label{ss2} You will pass in the {\it first} dialog box
"Parameters of the system". In this box you must set the
particle mass and stiffness coefficient in SI
units\footnote{\small In the line of the context-sensitive
help (bottom line of the display) {\it range of values} is
indicated within the bounds of {all parameters} numerical
values, you can change.} and write down those values to
table~1.

Let's note that the ending of input in all dialog boxes is
possible by two paths:
\begin{itemize}
\item[\ -]
by pressing the key \fbox{$Enter $};
\item[\ -]
by activation of the dialog box button $\fbox{Ok} $ (with the
help of the device $Mouse $)
\end{itemize}
\subsubsection{}
\label{ss3} In the following dialog box (according to its
title "Epoch angle") you should choose an epoch angle
$\varphi_o $ of a mass point oscillatory
motion\footnotemark[2]. Then write down this value to table~1
and press \fbox{$Enter $}~.
\subsubsection{}
\label{ss4} You will see the diagram representing a
trajectory of a free harmonic oscillatory motion of a mass
point with parameters chosen by you\footnote{\small If the
obtained diagrams  satisfy the object of the practice (in
your opinion) then sketch these diagrams in yours
writing-book and press any key (according to the message in
the bottom line).}.

\subsubsection{}
\label{ss5} You will pass at the next dialog box "Change of
mass". Enter another value of the particle mass (in
comparison with the value entered in item \ref{ss2} i.e. in
the dialog box "Parameters of the system" ). You will see
{\bf two} diagrams corresponding to different values of the
particle mass with unchangeable others parameters. By
pressing any key (according to the message in the bottom
line) you will return to the same dialog box again. Iterate
the described activities for one more value of mass. As a
result you will see {\bf\large three} diagrams corresponding
to {\it three} values of the particle mass is in a simple
harmonic motions\footnotemark[3].
\subsubsection{}
\label{ss6} You will pass at the next dialog box "Change of
K". Enter another value of the stiffness coefficient $k $ (in
comparison with value entered in item \ref{ss2} i.e. in the
dialog box "Parameters of the system" ). You will see {\bf
two} diagrams corresponding to different values of the
stiffness coefficient with unchangeable others parameters. By
pressing any key (according to the message in the bottom
line) you will return to the same dialog box again. Iterate
the described activities for one more stiffness coefficient.
In result you will see {\bf\large three} diagrams
corresponding to {\it three} values of the stiffness
coefficient of the quasi-elastic force (by due to action of
this force the mass point is in a simple harmonic
motions)\footnotemark[3].
\subsubsection{}
\label{ss7} You will pass at the next dialog box "Change of
epoch angle". Enter another value of the epoch angle (in
comparison with value entered in item \ref{ss3} i.e. in the
dialog box "Epoch angle" ). You will see {\bf two} diagrams
corresponding to different values of the epoch angle with
unchangeable others parameters. By pressing any key
(according to the message in the bottom line) you will return
to the same dialog box again. Iterate the described
activities for one more value of epoch angle. In result you
will see {\bf\large three} diagrams corresponding to {\it
three} values of the epoch angle of the simple harmonic
motions\footnotemark[3]. \ \\

So the first practice is ended and you will be returned to
the {\it main} menu. Let's note that {\it after you exit out
of the first practice you can} {\bf\large not} {\it enter
there} {\bf\large once again}. Therefore, if you do not
accept results of this practice and you want to iterate it
you should start the program once again.

\subsection{Damped oscillations}
\label{sb5}

In this practice state problem to study an oscillatory motion
of a mass point with the resistance force is proportional to
a vector of velocity. Namely:
\begin{itemize}
\item[\ -]
to make sure that the trajectory of a mass point is the non
harmonic function, but almost-periodic function;
\item[\ -]
to find out how the mass point trajectory varies by change of
the following parameters:
\begin{itemize}
\item[\ -]
particle mass $m $, stiffness coefficient $k\: $ in
expression for the elastic force (eq. (\ref{e3})) and the
resistance coefficient $b $ for the resistance force (eq.
(\ref{e6}))
\item[\ -]
initial kinematic parameters of a motion: the mass point
coordinates of the origin $x(0) = x_o \sin(\varphi_o) $ and
its initial velocity $v(0) = x_o \omega \cos(\varphi_o)\: $
(they are those parameters you can change by changing value
of the oscillations epoch angle $\varphi_o $);
\end{itemize}
\end{itemize}
\noindent The practice consists of the following items:
\subsubsection{}
\label{ss8} After you exited out of the first practice you
will see the main menu with titles of all practices again. By
keys \fbox{$\uparrow $} and \fbox{$\downarrow $} it is
necessary to select the practice "Damped oscillations" and
press \fbox{$Enter $}.
\subsubsection{}
\label{ss9} You will pass in the {\it first} dialog box
"Parameters of the system". In this box you must set the
particle mass, stiffness coefficient and resistance
coefficient in SI units\footnotemark[2] and write down those
values to table~2.
\subsubsection{}
\label{ss10} In the following dialog box (according to its
title "Epoch angle") you should choose an epoch angle
$\varphi_o $ of mass point oscillatory
motion\footnotemark[2]. Then write down this value to table~2
and press \fbox{$Enter $}~.
\subsubsection{}
\label{ss11} You will see the diagram representing a
trajectory of a damped oscillatory motion of a mass point
with parameters chosen by you\footnotemark[3].
\subsubsection{}
\label{ss12} You will pass at the next dialog box "Change of
the parameters". Here you can change values of two
parameters: the particle mass $m $ and resistance coefficient
$b $. In contrast to the first practice, {\it you can have
this box as much as long}. Because {\it after each new
diagram (for the next pair of parameters m and b) you will be
returned here}. However, as well as in the first practice, at
the display draw no more three diagrams. Therefore we
recommend to act as follows:
\begin{itemize}
\item[\ -]
at first draw {\bf\large three} diagrams with {\it different
values of the particle mass} $m $ and the {\it constant}
resistance coefficient $b $\footnotemark[3];
\item[\ -]
then draw {\bf\large three} diagrams with {\it different
values of the resistance coefficient} $b $ and the {\it
constant} particle mass $m $\footnotemark[3];
\item[\ -]
then select the such {\it underload} resistance coefficient
$b_{min} $ (with the constant particle mass $m $) for which
the particle motion will become aperiodic; write down values
$m $ and $b_{min} $ to table~2;
\item[\ -]
iterate operations of the previous item for {\bf another two}
values of the particle mass $m $;
\item[\ -]
then enter pairwise obtained values of mass $m $ and
coefficient $b_{min} $ (from table~2) so to obtain {\it all
three} aperiodic motions on one picture (i.e. in one frame)
and {\bf\Large sketch} these diagrams in your writing-book;
\end{itemize}

So the second practice is ended. In order to return to the
{\it main} menu (if you have the dialog box "Change of
parameters") is necessary to press the key \fbox{Esc} (or
make active the dialog box button $\fbox{Exit} $ by the
device $Mouse $).

\section{Return\protect\footnote{\small{\bf Title
page} of the return at the laboratory work on physical
processes simulation one should draw up on the same rules
that the title page of the return at the laboratory work is
done in chair T\&EPh experimental laboratories.}}

\subsection{Contents of the return}

\noindent The return should include the following items:
\begin{itemize}
\item[\ 1.] {\it Object of work.}
\item[\ 2.] {\it Summary theoretical part.}
\item[\ 3.] {\it Practices:}
\begin{itemize}
\item[\ - ] {\bf Free simple harmonic motions.}\\
 T A B L E~~1\\
Diagrams \fbox{~$x = x(t)$~~with different $m $~} for all
three trajectories in one frame.\\ Diagrams \fbox{~$x =
x(t)$~~with different $k $~~} for all three trajectories in
one frame.\\ Diagrams \fbox{~$x = x(t)$~~with different
$\varphi_o $~} for all three trajectories in one frame.\\
\item[\ - ] {\bf Damped oscillations.}\\
 T A B L E~~2\\
Diagrams \fbox{~$x = x(t)$~~with different $m $~~~} for all
three trajectories in one frame.\\ Diagrams \fbox{~$x =
x(t)$~~with different $b\, $~~~~} for all three trajectories
in one frame.\\ Diagrams \fbox{~$x = x(t)$~~with different
$b_{min} $} for all three trajectories in one frame.\\
\end{itemize}
\item[\ 4.] {\it Conclusion}
\end{itemize}

\subsection{Design of the tables.}

\noindent The tables used in the return should be designed by
following ways.

\begin{center}
\begin{tabular}{||l|c|c|c||}
\multicolumn{4}{r}{ T A B L E  1}\\ \hline
 &
 Change $m $ &
 Change $k $ &
 Change $\varphi_o $ \\
\hline \hline
 $m,\:  $ &
 {\hfill \vline} {\hfill \vline} {\hfill } &
 &
 \\
\hline
 $k,\:  $ &
 &
 {\hfill \vline} {\hfill \vline} {\hfill } &
 \\
\hline
 $\varphi_o,\:  $ &
 &
 &
 {\hfill \vline} {\hfill \vline} {\hfill } \\
\hline \hline
\end{tabular}

\ \\

\ \\

\begin{tabular}{||l|c|c|c||}
\multicolumn{4}{r}{ T A B L E  2}\\ \hline
 &
 Change $m $ &
 Change $b $ &
 Change $b_{min} $ \\
\hline \hline
 $m,\:  $ &
 {\hfill \vline} {\hfill \vline} {\hfill } &
 &
 {\hfill \vline} {\hfill \vline} {\hfill } \\
\hline
 $b,\:  $ &
 &
 {\hfill \vline} {\hfill \vline} {\hfill } &
 \\
\hline
 $b_{min},\: $ &
 &
 &
 {\hfill \vline} {\hfill \vline} {\hfill } \\
\hline
 $k,\:  $ &
 \multicolumn{3}{|c||}{} \\
\hline
 $\varphi_o,\:  $  &
 \multicolumn{3}{|c||}{} \\
\hline \hline
\end{tabular}
\end{center}

\section{Conclusion}
Let's mark, that this paper is written on the
basis of the previous works \cite{pp1} carried out on chair T\&EPh under
the author leadership (or direct participation).

\end{document}